\documentclass{emulateapj_mo}
\usepackage{lscape}

\shorttitle{SDSS J1029+2623: 22 arcsecond lensed QSO}
\shortauthors{Inada et al.}

\begin{document}

\title{SDSS J1029+2623: A Gravitationally Lensed Quasar with an Image
  Separation of 22.5 Arcseconds${}^1$} 

\author{
Naohisa Inada\altaffilmark{2,3},
Masamune Oguri\altaffilmark{4,5},
Tomoki Morokuma\altaffilmark{2},
Mamoru Doi\altaffilmark{2},
Naoki Yasuda\altaffilmark{6},
Robert H. Becker\altaffilmark{7,8},
Gordon T. Richards\altaffilmark{9,10}, 
Christopher S. Kochanek\altaffilmark{11},
Issha Kayo\altaffilmark{12}, 
Kohki Konishi\altaffilmark{6},
Hiroyuki Utsunomiya\altaffilmark{2},
Min-Su Shin\altaffilmark{5}, 
Michael A. Strauss\altaffilmark{5}, 
Erin S. Sheldon\altaffilmark{13},
Donald G. York\altaffilmark{14,15},
Joseph F. Hennawi\altaffilmark{16}, 
Donald P. Schneider\altaffilmark{17},
Xinyu Dai\altaffilmark{11},
and 
Masataka Fukugita\altaffilmark{6}
}
\altaffiltext{1}{Based in part on data collected at Subaru Telescope, 
    which is operated by the National Astronomical Observatory of Japan.}
\altaffiltext{2}{Institute of Astronomy, Faculty of Science, University
of Tokyo, 2-21-1 Osawa, Mitaka, Tokyo 181-0015, Japan.} 
\altaffiltext{3}{Japan Society for the Promotion of Science (JSPS)
  Research Fellow.} 
\altaffiltext{4}{Kavli Institute for Particle Astrophysics and
  Cosmology, Stanford University, 2575 Sand Hill Road, Menlo Park, CA
  94025.} 
\altaffiltext{5}{Princeton University Observatory, Peyton Hall,
  Princeton, NJ 08544.}  
\altaffiltext{6}{Institute for Cosmic Ray Resarch, University of Tokyo, 
5-1-5 Kashiwa, Kashiwa, Chiba 277-8582, Japan.} 
\altaffiltext{7}{IGPP-LLNL, L-413, 7000 East Avenue, Livermore, CA 94550.}
\altaffiltext{8}{Department of Physics, University of California at
  Davis, 1 Shields Avenue, Davis, CA 95616.}  
\altaffiltext{9}{Department of Physics, Drexel University, 3141
  Chestnut Street,  Philadelphia, PA 19104.}
\altaffiltext{10}{Johns Hopkins University, 3400 N. Charles St.,
  Baltimore, MD 21218.} 
\altaffiltext{11}{Department of Astronomy, The Ohio State University, 
  Columbus, OH 43210.}
\altaffiltext{12}{Department of Physics and Astrophysics, Nagoya
University, Chikusa-ku, Nagoya 464-8062, Japan.} 
\altaffiltext{13}{Center for Cosmology and Particle Physics, 
Department of Physics, New York University, 4 Washington Place, New
York, NY 10003. }   
\altaffiltext{14}{Department of Astronomy and Astrophysics, The
  University of Chicago,  5640 South Ellis Avenue, Chicago, IL 60637.}
\altaffiltext{15}{Enrico Fermi Institute, The University of Chicago,
  5640 South Ellis Avenue, Chicago, IL 60637.}
\altaffiltext{16}{Department of Astronomy, University of California at
  Berkeley, 601 Campbell Hall, Berkeley, CA 94720-3411.}  
\altaffiltext{17}{Department of Astronomy and Astrophysics, The
  Pennsylvania State University, 525 Davey Laboratory, University
  Park, PA 16802.}   
 
\setcounter{footnote}{17}

\begin{abstract}
We report the discovery of a cluster-scale lensed quasar, 
SDSS J1029+2623, selected from the Sloan Digital Sky Survey. The lens
system exhibits two lensed images of a quasar at ${z_s}=2.197$. The
image separation of $22\farcs5$ makes it the largest separation lensed
quasar discovered to date. The similarity of the optical spectra and 
the radio loudnesses of the two components support the
lensing hypothesis. Images of the field show a cluster of galaxies at
${z_l}\sim 0.55$ that is responsible for the large image
separation. The lensed images and the cluster light center are not
collinear, which implies that the lensing cluster has a complex
structure.  
\end{abstract}


\section{Introduction}\label{sec:intro}

The discovery of SDSS J1004+4112 with an image separation of 
$14\farcs6$, the first example of a quasar multiply imaged by a massive
cluster of galaxies, opened a new window for understanding our 
universe \citep{inada03,oguri04a,sharon05}. 
Although there are many examples of galaxies (arcs) lensed by 
clusters,
large-separation lensed quasars have several advantages over arcs 
as a cosmological probe. 
First, the simpler (point-like) structure of quasars and
their well-understood redshift distribution should make the 
large-separation lensed quasars much cleaner probes of cosmology and 
structure formation models \citep[e.g.,][]{oguri04b,hennawi06}, while 
the statistics of arcs remain contentious \citep{bartelmann03}. 
Second, the time-variability of quasars allows the measurement of 
time delays among the multiple lensed images, thereby breaking 
the mass-sheet degeneracy of lens models given a priori knowledge 
of the Hubble constant \citep[e.g.,][]{kochanek02}. 
This was explored in detail for SDSS J1004+4112 
\citep[e.g.,][]{oguri04a,williams04,kawano06,fohlmeister06}. 
Our current problem is that the small number of known systems 
limits their use in statistical analyses, however, 
current and future large lens surveys will discover
many large-separation lensed quasars \citep{wambsganss03}.

In this {\it Letter}, 
we report the discovery of the second large-separation lensed quasar, 
SDSS J1029+2623, a quasar at $z_s=2.197$ doubly imaged by a massive galaxy  
cluster at $z_l{\sim}0.55$. 
It was discovered in the course of the Sloan Digital
Sky Survey Quasar Lens Search \citep[SQLS;][]{oguri06}, which is a
survey of strongly lensed quasars in
the Sloan Digital Sky Survey \citep[SDSS;][]{york00}.
The image separation of $22\farcs5$ makes it 
the largest separation lens among the $\sim 100$ lensed quasars known 
so far \citep{kochanek06}.

\section{Candidate Selection and Subaru observations}\label{sec:sdss}

\begin{figure*}
\epsscale{.32}
\plotone{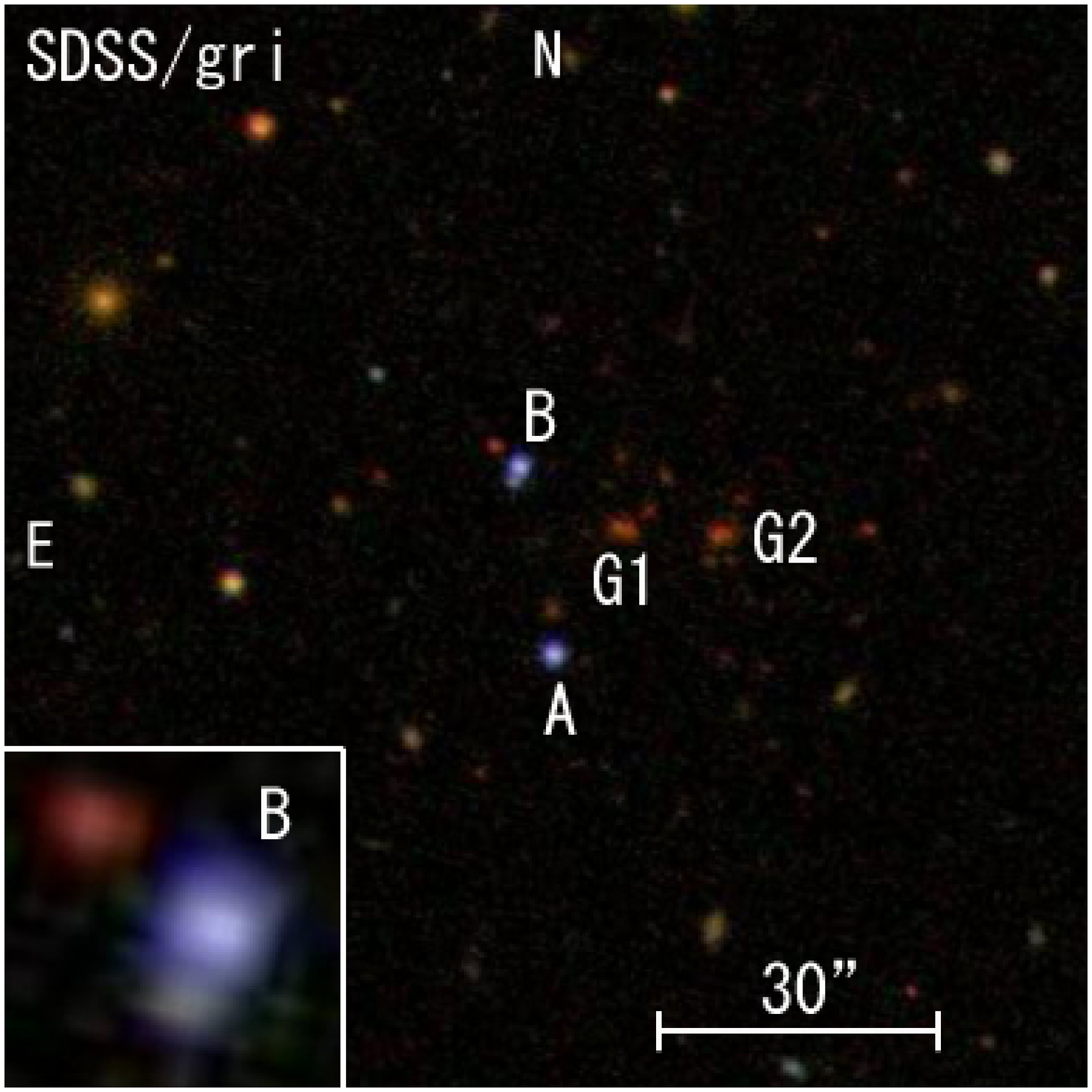}
\plotone{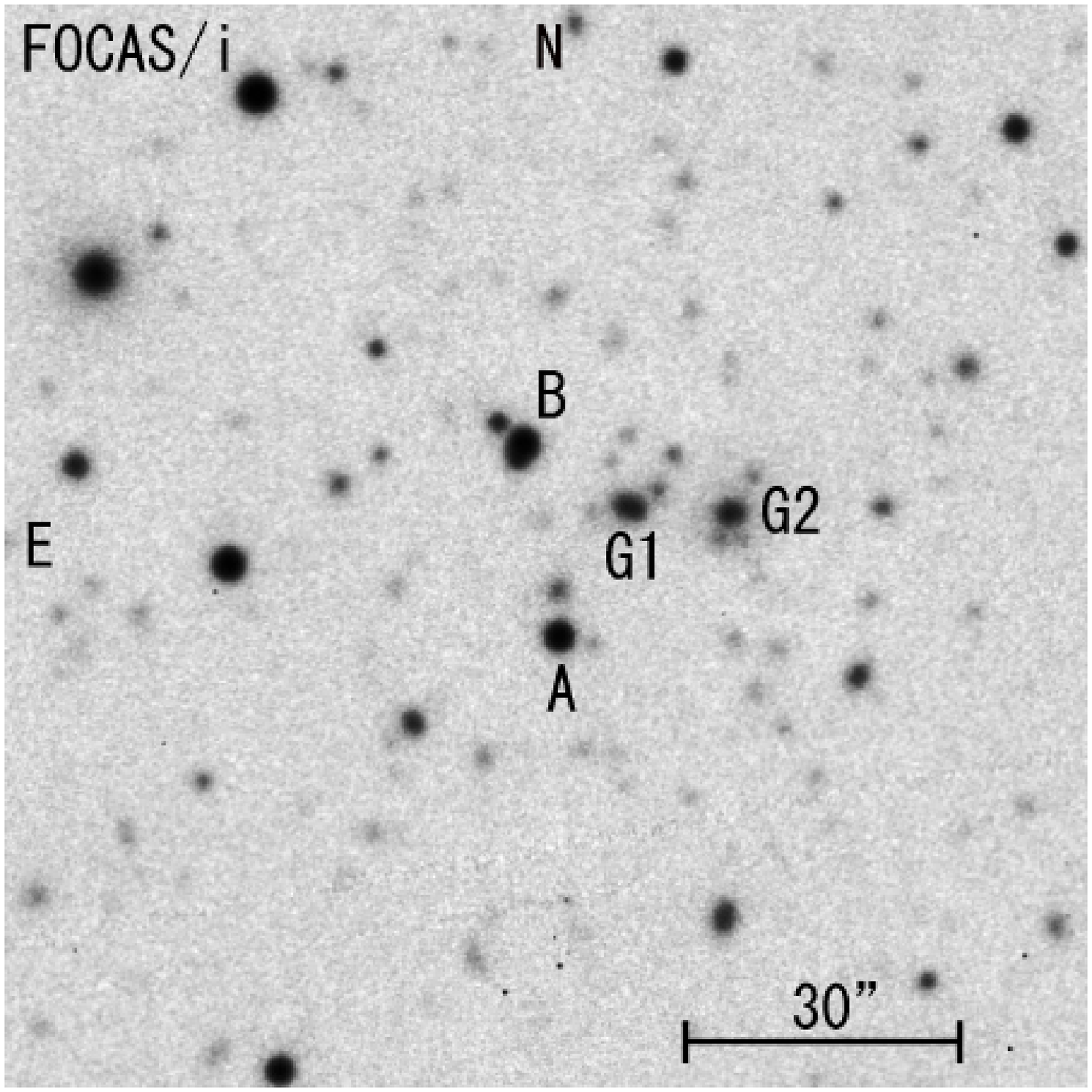}
\plotone{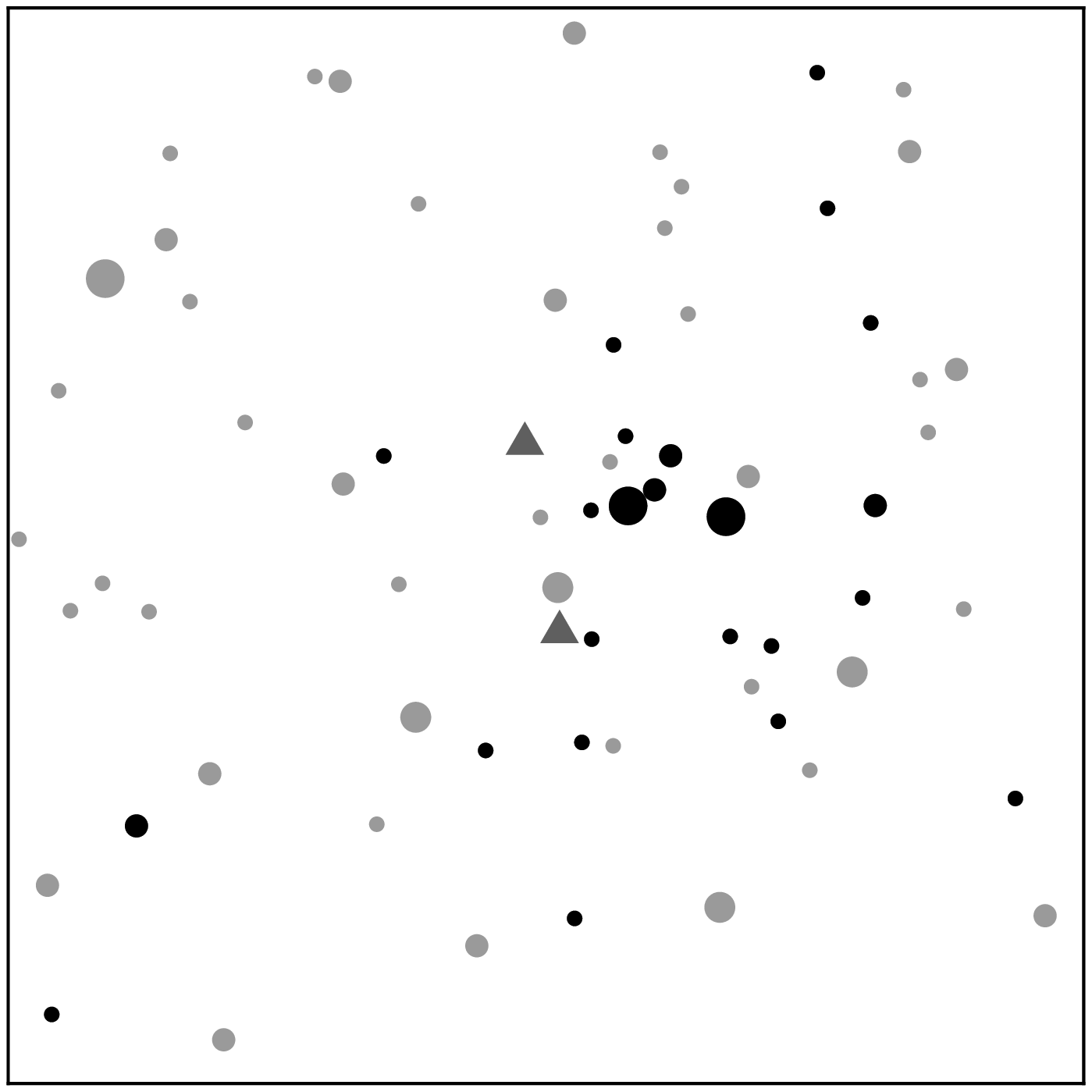}
\caption{{\it Left}: The $gri$ composite SDSS image of SDSS J1029+2623 
 (1\farcs3 seeing). The quasar images (blue stellar objects) are indicated 
 by A and B. G1 and G2 (red extended objects) are likely to be member 
 galaxies of a lensing cluster at $z{\sim}0.55$. 
 The inset shows an expanded view of component B: An object $\sim 2''$
 Southeast of component B has quite different color from those of the 
 quasar components. 
 {\it Middle}: The FOCAS $i$-band image (1\farcs8 seeing).
 {\it Right}: Locations of galaxies ($i<22.5$) identified in the
  FOCAS images are plotted with filled circles. Larger circles mean
  brighter galaxies. Galaxies that survive the color cut of
  $R-i=0.7{\pm}0.3$ are shown by darker circles. Filled triangles are
  the locations of the two quasar images. In all the panels, 
  North is up and East is left. 
\label{fig:img}}
\end{figure*}

From the SDSS-II Sloan Legacy Survey data, we discovered the 
lens candidate SDSS J1029+2623 
using the algorithm described in \citet{oguri06}.
It consists of a quasar whose redshift was spectroscopically 
measured by the SDSS to be $z=2.198$ (hereafter referred to as A)
and a stellar object (hereafter B) with almost 
the same color as component A (see Table \ref{tab:info}). 
Components A and B are separated by $22\farcs5$. 
There are no other point sources in the field with similar 
colors that would be candidates for additional quasar images.
In addition to the 
similar colors, the $gri$ composite SDSS image shown in the left panel of
Figure \ref{fig:img} exhibits a concentration of red galaxies centered on 
two bright galaxies (named G1 and G2), which implies the existence 
of a high-redshift cluster of galaxies. Indeed, the photometric 
redshifts of G1 and G2 are both $\sim$0.55 from their SDSS colors 
\citep{csabai03}. The properties of components A, B, G1, and G2 
are summarized in Table \ref{tab:info}. 
We note that the details of the SDSS are described in a series 
of technical papers: \citet{gunn06} for the dedicated wide-field 
($3^{\circ}$ field of view) 2.5-m telescope; 
\citet{fukugita96}, \citet{gunn98}, \citet{lupton99}, \citet{hogg01},
\citet{lupton01}, \citet{smith02}, \citet{pier03}, \citet{ivezic04}, 
and \citet{tucker06} for the photometric survey; 
\citet{richards02} for the spectroscopic target selection algorithm 
of quasars; and \citet{blanton03} for the tiling algorithm of 
the spectroscopic survey. Most of the SDSS data are already publicly 
available \citep{stoughton02,abazajian03,abazajian04,abazajian05,adelman06}.    

\begin{figure}
\includegraphics[scale=0.35,angle=270]{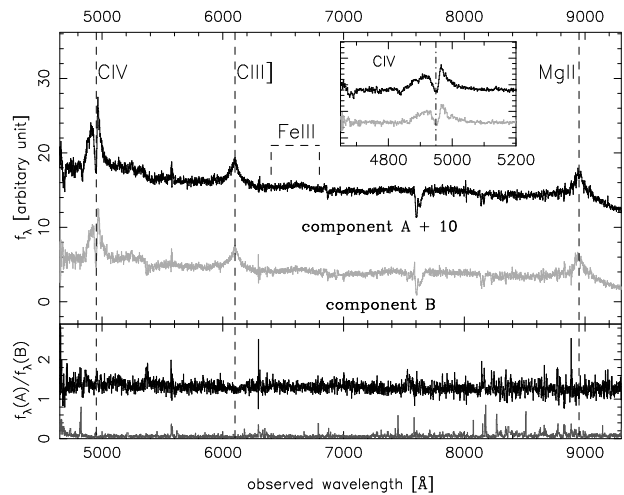}
\caption{Spectra of components A ({\sl black} solid line, 
  shifted upwards by a constant of 10 flux density units) 
  and B ({\sl gray} solid line) taken with the FOCAS on the 
  Subaru telescope. The signal to noise ratio 
  (in the 6000{\,\AA} -- 8000{\,\AA} continuum) is about 25.0 per pixel. 
  The dotted lines indicate the positions of the
  quasar \ion{C}{4}, \ion{C}{3]}, \ion{Fe}{3}, and \ion{Mg}{2}
  emission lines redshifted to $z=2.197$. The feature at 7600{\,\AA}
  is atmospheric. The inset shows an expanded view of the \ion{C}{4}
  emission lines. The vertical dot-dashed line in the inset shows the
  central position of the broad absorption lines. 
  In the bottom panel, the black solid
  line shows the spectral flux ratio between A and B, and the dark
  gray solid line shows the 1$\sigma$ error of the spectral flux
  ratio, which is derived from the noise per pixel in each
  spectrum. The ratio is confirmed to be constant (${\sim}1.2$) within
  the $2.3{\sigma}$ error, over the entire range of observed
  wavelengths. 
\label{fig:spec}}
\end{figure}

We obtained 600~sec spectra of components A and B with the Faint
Object Camera And Spectrograph \citep[FOCAS;][]{kashikawa02} at the
Subaru 8.2-meter telescope on 2006 June 28. The observation was
conducted in the 3$\times$1 on-chip binning mode, using the 300B grism, 
the SY47 filter and a $0\farcs8$-width slit aligned along components A and B. 
The spectral resolution, wavelength coverage, and spatial scale 
of the CCD detector were $\hbox{R}\sim500$, from 4700{\,\AA} to
9400{\,\AA}, and $0\farcs311$ pixel$^{-1}$, respectively. 
Although the seeing was poor ($\sim$1\farcs8), the $22\farcs5$ separation
makes it easy to extract the spectra of the two components using 
standard IRAF\footnote{
  IRAF is distributed by the National Optical Astronomy Observatories,
   which are operated by the Association of Universities for Research
   in Astronomy, Inc., under cooperative agreement with the National
   Science Foundation.} 
tasks. 
The spectra are shown in the upper panel of Figure \ref{fig:spec}. 
It is clear that both components are quasars at the same
redshift, $z=2.197$ (see Table \ref{tab:info}). Moreover, the
spectral shapes are similar; both components have similar 
broad absorption line features at the same wavelength in the \ion{C}{4} 
emission lines, similar profiles for the \ion{Fe}{3} emission lines,
and similar characteristic shapes of the red wings of the \ion{Mg}{2} 
emission lines. In addition, the spectral flux ratio plotted in the 
lower panel of Figure \ref{fig:spec} is constant (${\sim}1.2$) 
over the full range of the observed wavelengths. The similarity of the 
spectra supports the idea that the two components are 
lensed images of a single quasar.

We also obtained 120~sec $R$- and 120~sec $i$-band images 
($\sim$1\farcs8 seeing)  
with FOCAS. On-chip $2\times2$ binning yielded images with a
pixel scale of 0\farcs207 pixel$^{-1}$. The $i$-band image is
shown in the middle panel of Figure \ref{fig:img}. From the FOCAS images, 
we found that galaxy G1 is likely to be a superposition of two
early-type galaxies; however, we treated them as a single object since
the poor seeing prevents us from separating them. We used SExtractor
\citep{bertin96} for the photometry (MAG\_AUTO) calibrated by the SDSS
magnitudes of the nearby stars\footnote{
  We used the transformation
  formulae described in Table 7 of \citet{smith02} to calculate $R$
  magnitudes of the nearby stars from the SDSS $gr$ magnitudes.}.
All galaxies (classified by the combination of the SExtractor 
CLASS\_STAR parameter and visual inspection) with $i<22.5$ are 
plotted as filled circles in the right panel of Figure
\ref{fig:img}. Among them, we selected galaxies with colors of
$R-i=0.7{\pm}0.3$ as possible cluster members, because G1 and G2 have
$R-i$ colors of 0.6 and 0.8, respectively. The $R-i$ color of 0.7 is
consistent with the red-sequence color of a galaxy cluster at
$z{\sim}0.55$ \citep[e.g.,][]{goto03}. The selected galaxies are
plotted as dark filled circles in the right panel of Figure
\ref{fig:img}. As expected, clustering of galaxies 
near G1 and G2 is much more pronounced after the color cut, 
suggesting the existence of a cluster of galaxies at 
$z\sim0.55$. 

\section{Discussion and Conclusion}\label{sec:subaru}

\begin{figure}
\epsscale{.85}
\plotone{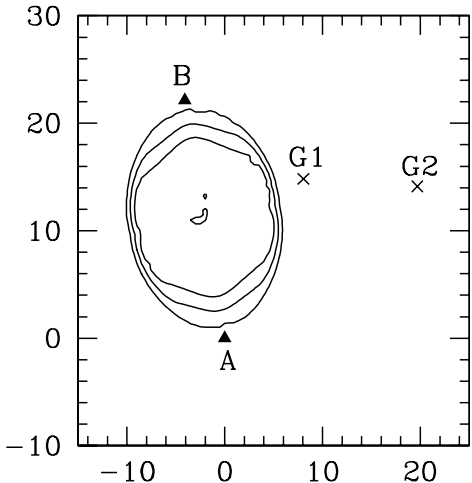}
\caption{Constraints on the center of the potential of the lensing cluster 
  from mass modeling. Solid lines indicate 1, 2, and $3\sigma$
  likelihood regions for the center based on an SIE model for the cluster 
  potential (see text for details). 
  The quasar images (A and B) and bright member galaxies 
  (G1 and G2) are marked by filled triangles and crosses, respectively. 
  If we model the system using a SIE+external shear with the potential 
  center fixed to galaxy G1, the best-fitting model has an Einstein radius
  of $15\farcs0$ (corresponding to the velocity dispersion of 
  $\sim$900 km sec$^{-1}$), an ellipticity of 0.25 with its direction 
  of $88^\circ$ (East of North), and a shear of 0.09 with its direction 
  of $28^\circ$, although it does not reproduce the observables exactly. 
\label{fig:model}}
\end{figure}

An interesting thing about this lens system is that the quasar
images and the light center of the lensing cluster appear not to be
collinear (see Figure \ref{fig:img}). 
Brightest cluster galaxies are sometimes significantly offset from the 
center of a cluster mass \citep[e.g.,][]{lin04}, so the lack of 
collinearity does not necessarily argue against the lensing hypothesis. 
Nevertheless, it is of great interest to understand the impact of this 
offset for lens models. To explore this, we use standard mass
modeling techniques as implemented in the public {\it lensmodel}
software \citep{keeton01}. Since the small number of observational 
constraints limits detailed investigations of the mass distribution, 
we adopt the simplest model used for modeling studies, a singular
isothermal ellipsoid (SIE). We fit the quasar positions 
and fluxes
assuming a position error of $0\farcs5$ and a flux error of 20\%; 
we adopt errors larger than the actual measurement errors in order to
account for perturbations from massive galaxies in the cluster. 
In addition we include a mild prior on the ellipticity, $e=0.3\pm0.2$,
to exclude models with a highly elongated cluster. With this model,
we derive the allowed range for the center position of the lensing
cluster potential, as shown in Figure \ref{fig:model}. 
The offset of at least $\sim 30$~kpc between the preferred model
positions and galaxies G1 and G2 suggests either that the actual mass
distribution of the lensing cluster is more complex than this 
simple model or that there is a genuine offset that might be confirmed 
in X-ray observations.

\begin{figure}
\epsscale{.9}
\plotone{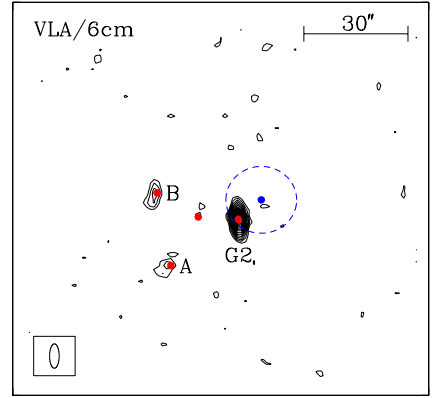}
\caption{The 6 cm VLA radio map of SDSS J1029+2623 field.
 North is up and East is left. 
 The observation was made during the transition between the B and
 CnB arrays, leading to the strongly elongated beam (shown in the 
 lower left inset). The red small dots 
 represent the optical position of components A, B, G1, and G2. 
 The small blue dot Northwest of G2 marks the ROSAT X-ray source position 
 while the surrounding blue dashed circle indicates the uncertainties in 
 the position.
\label{fig:vla}}
\end{figure}

While the similarity of the optical spectra and the existence of the 
lensing galaxy cluster
make the lensing interpretation solid, we can further test the lensing
hypothesis using archival data at other wavelengths.  
In the FIRST radio survey \citep{becker95}, there are two radio
sources in the field: Galaxy G2, with a peak flux of 4.20 mJy/beam, 
is the strongest source, component B, with a peak flux of 1.01 mJy/beam, 
is detected just at the flux limit, and component A is not detected.
Although the difference 
could be explained by quasar variability combined with the time 
delay between the lensed images, it might argue a binary interpretation 
for the quasars. 
In order to clarify the interpretation of the FIRST radio data, 
we obtained a 1 hour 6 cm Very Large Array (VLA) radio map of the system 
on 23 September 2006.  
Figure \ref{fig:vla} shows the primary beam corrected VLA 6 cm map; 
both quasar components have more than
5$\sigma$ detections (0.237$\pm$0.045 mJy/beam for A and 
0.325$\pm$0.045 mJy/beam for B). There are some offsets 
(but within 5$\sigma$, see Table \ref{tab:info}) between the 
SDSS positions and the VLA positions of the quasar components.
However both components are extended in the VLA data and 
better quality data will be necessary to clarify the offsets.
The radio loudnesses of the images, defined relative to the SDSS 
$i$-band fluxes, are 1.8 and 2.5 for images A and B, respectively.  
The radio flux ratio (A/B) 
of 0.73 is mildly inconsistent with the optical flux ratio 
of 1.2, however, allowing for variability both between the epochs 
of the two observations and the multi-year time delay of 
the two images, finding two quasars with such similar radio loudnesses 
further supports the lensing hypothesis given the enormous range of 
radio loudnesses observed for quasars \citep{ivezic02}.

The system was also detected as 1RXS J102912.0+262338 by the ROSAT
All-Sky Survey \citep{voges99} with a count rate of
$0.040 \pm 0.016$ counts s${}^{-1}$. 
This corresponds to an 0.1-2.4~keV flux of approximately 
$(4\pm 1)\times 10^{-13}$ ergs cm$^{-2}$ s$^{-1}$, and
it is flagged as an extended source, albeit at low significance. 
The X-ray position
could be consistent with the position of G2 given the
$10\farcs0$ uncertainties in the X-ray data (Figure \ref{fig:vla}). 
Assuming the X-ray signals arise from the lensing cluster ($z{\sim}0.55$),
the X-ray luminosity is approximately $3 \times 10^{44} h^{-2}$ ergs s${}^{-1}$
(for a cosmological model with $\Omega_M=0.3$ and $\Omega_\Lambda=0.7$)
and corresponds to a velocity dispersion of approximately 
$\sigma \simeq 1000$~km s$^{-1}$ that is slightly larger than 
the velocity scale of $\sigma \simeq 900$~km s$^{-1}$ needed
to produce the 22\farcs5 image separation. Thus the
X-ray luminosity is a little too high, suggesting that the
AGN in galaxy G2 is an X-ray source as well as a radio source, 
even though it shows no signs of AGN activity in its broad band
optical colors. 

In summary, the best interpretation of the two quasar components 
in the SDSS J1029+2623 system is that a quasar at ${z_s}=2.197$ is doubly 
imaged by a cluster of galaxies at ${z_l}\sim 0.55$. 
The evidence for strong lensing comes from 
the remarkable similarity of the spectral shapes 
and the similar radio loudnesses of the two quasars, and 
the existence of the lensing cluster with the presence of an 
extended X-ray source which is capable of producing the observed 
image separation. However, simple mass models require a center of the 
cluster potential offset from the positions of the bright galaxies or 
the ROSAT X-ray emission. Such an offset with respect to the bright galaxies 
was also found for the SDSS1004+4112 system \citep{oguri04a}, 
and the X-ray position may be dominated by the AGN in galaxy G2.
Thus, further follow-up observations, such as deep and 
high-resolution X-ray imaging and 
optical spectroscopy of member galaxies are needed
to identify the cluster potential center. 

\acknowledgments

This work was supported in part by Department of Energy contract
DE-AC02-76SF00515.
A portion of this work 
was also performed under the auspices of the U.S. Department of Energy,
National Nuclear Security Administration by the University of
California, Lawrence Livermore National Laboratory under contract
No. W-7405-Eng-48. 
The National Radio Astronomy Observatory is a facility of the National 
Science Foundation operated under cooperative agreement by Associated 
Universities, Inc. This work made use of the Very Large Array at the 
NRAO, and we thank NRAO for a rapid response time award.

Funding for the SDSS and SDSS-II has been provided by the 
Alfred P. Sloan Foundation, the Participating Institutions, 
the National Science Foundation, the U.S. Department of Energy, 
the National Aeronautics and Space Administration, the Japanese 
Monbukagakusho, the Max Planck Society, and the Higher Education 
Funding Council for England. The SDSS Web Site is http://www.sdss.org/.

The SDSS is managed by the Astrophysical Research Consortium for 
the Participating Institutions. The Participating Institutions 
are the American Museum of Natural History, Astrophysical Institute 
Potsdam, University of Basel, Cambridge University, Case Western 
Reserve University, University of Chicago, Drexel University, 
Fermilab, the Institute for Advanced Study, the Japan Participation 
Group, Johns Hopkins University, the Joint Institute for Nuclear 
Astrophysics, the Kavli Institute for Particle Astrophysics and 
Cosmology, the Korean Scientist Group, the Chinese Academy of 
Sciences (LAMOST), Los Alamos National Laboratory, the Max-Planck-Institute 
for Astronomy (MPIA), the Max-Planck-Institute for Astrophysics (MPA), 
New Mexico State University, Ohio State University, University of 
Pittsburgh, University of Portsmouth, Princeton University, the 
United States Naval Observatory, and the University of Washington.

\clearpage
\begin{landscape}
\begin{deluxetable}{ccccccccccc}
\tablecaption{Properties of SDSS J1029+2623\label{tab:info}}
\tablewidth{0pt}
\tablehead{
 \colhead{Component}   & \colhead{R.A.(SDSS)} &
 \colhead{Decl.(SDSS)} & \colhead{R.A.(VLA)} &
 \colhead{Decl.(VLA)}  & \colhead{$i$} & 
 \colhead{$u-g$} & \colhead{$g-r$} 
 & \colhead{$r-i$} & \colhead{$i-z$} 
 & \colhead{Redshift\tablenotemark{a}}
}
\startdata
A  & 10:29:13.94 & +26:23:17.9 & 10:29:14.05\tablenotemark{b} & +26:23:17.4\tablenotemark{b} & 18.59$\pm$0.01 & 0.72$\pm$0.04 & 0.18$\pm$0.01 & 0.19$\pm$0.01 & 0.32$\pm$0.03 & 2.1966$\pm$0.0003 \\
B  & 10:29:14.24 & +26:23:40.1 & 10:29:14.31\tablenotemark{b} & +26:23:39.3\tablenotemark{b} & 18.61$\pm$0.01 & 0.67$\pm$0.04 & 0.14$\pm$0.01 & 0.20$\pm$0.01 & $-0.10$$\pm$0.04 & 2.1969$\pm$0.0003 \\
G1 & 10:29:13.35 & +26:23:32.8 & \nodata\phn & \nodata\phn & 19.01$\pm$0.04 & 0.71$\pm$0.90 & 1.93$\pm$0.17 & 1.05$\pm$0.06 & 0.75$\pm$0.09 & 0.56$\pm$0.03 \\
G2 & 10:29:12.48 & +26:23:32.0 & 10:29:12.49 & +26:23:31.9 & 18.77$\pm$0.03 & 1.63$\pm$1.72 & 1.87$\pm$0.15 & 1.06$\pm$0.05 & 0.48$\pm$0.08 & 0.53$\pm$0.06 \\
\enddata
\tablecomments{Data are from the SDSS except for the celestial coordinates in the 
  VLA data and the spectroscopic
  redshifts. Magnitudes refer model magnitudes without Galactic
  extinction corrections, measured on 13 December 2004.
  All celestial coordinates are given in J2000.}
\tablenotetext{a}{Spectroscopic redshifts (derived from \ion{Mg}{2}
emission lines in the FOCAS spectra) for A and B, and photometric
redshifts for G1 and G2 \citep{csabai03}.}  
\tablenotetext{b}{VLA position errors for A and B are both about 0\farcs5 
   per coordinate.} 
\end{deluxetable}
\clearpage
\end{landscape}

\end{document}